\documentclass{article}
\usepackage[margin=0.8in]{geometry}
\usepackage{amsmath,amsthm,amssymb,amsfonts,bm, xcolor}
\usepackage{hyperref,url,graphicx,multirow}
\newcommand{\N}{\ensuremath{\mathbb{N}}}
\newcommand{\hide}[1]{}
\usepackage{authblk}

\usepackage{algorithm,algpseudocode}
\usepackage[sorting = none, backend = bibtex, style=numeric-comp]{biblatex}
\title{Solving Larger Maximum Clique Problems Using Parallel Quantum Annealing}

\author[1]{Elijah Pelofske\thanks{Email: epelofske@lanl.gov}}
\author[2]{Georg Hahn}
\author[1,3]{Hristo N.\ Djidjev}
\affil[1]{Los Alamos National Laboratory, CCS-3, Los Alamos, NM 87545, USA}
\affil[2]{Harvard T.H.\ Chan School of Public Health, Boston, MA 02115, USA}
\affil[3]{Institute of Information and Communication Technologies, Bulgarian Academy of Sciences, Sofia, Bulgaria}

\date{\vspace{-6ex}}

\begin{document}
\maketitle

\begin{abstract}
Quantum annealing has the potential to find low energy solutions of NP-hard problems that can be expressed as quadratic unconstrained binary optimization problems. However, the hardware of the quantum annealer manufactured by D-Wave Systems, which we consider in this work, is sparsely connected and moderately sized (on the order of thousands of qubits), thus necessitating a minor-embedding of a logical problem onto the physical qubit hardware. The combination of relatively small hardware sizes and the necessity of a minor-embedding can mean that solving large optimization problems is not  possible on current quantum annealers. In this research, we show that a hybrid approach combining parallel quantum annealing with graph decomposition allows one to solve larger optimization problem accurately. We apply the approach on the Maximum Clique problem on graphs with up to 120 nodes and 6395 edges.
\end{abstract}

\section{Introduction}
\label{sec:introduction}
Quantum annealers are devices that aim to use quantum mechanical fluctuations to search for low energy solutions to combinatorial optimization problems \cite{hauke2020perspectives, morita2008mathematical, Kadowaki_1998, das2008colloquium}. Quantum annealers have been physically manufactured, such as those manufactured by D-Wave Systems, Inc.~\cite{johnson2011quantum, TechnicalDescriptionDwave, PhysRevE.58.5355, PhysRevX.4.021041, boixo2013experimental}, and allow approximate solutions of NP-hard problems to be computed that are hard to solve classically. D-Wave quantum annealers are designed to find high quality solutions of so-called quadratic unconstrained binary optimization (QUBO) and Ising problems by minimizing the function
\begin{align}
    H(x_1,\ldots,x_n) = \sum_{i=1}^n a_i x_i + \sum_{i<j} a_{ij} x_i x_j,
    \label{eq:hamiltonian}
\end{align}
where the linear weights $a_i \in \mathbb{R}$, $i \in \{1,\ldots,n\}$, and the quadratic couplers $a_{ij} \in \mathbb{R}$ for $i<j$ are chosen by the user to define the problem under investigation. The variables $x_i$, $i \in \{1,\ldots,n\}$, are binary and unknown. The function of eq.~\eqref{eq:hamiltonian} is called a QUBO if $x_i \in \{0,1\}$, and an Ising problem if $x_i \in \{-1,+1\}$ for all $i \in \{1,\ldots,n\}$. Both QUBO and Ising formulations are equivalent \cite{Lucas2014, Chapuis2017}.

After mapping eq.~\eqref{eq:hamiltonian} to the physical quantum system, the D-Wave quantum annealer aims to find the values of $x_1,\ldots,x_n$ that minimize $H$ by trying to obtain a minimum-energy state of the quantum system. To this end, the coefficients of the linear terms in eq.~\eqref{eq:hamiltonian} are mapped onto the corresponding qubits' physical parameters, and the coefficients for the quadratic terms are mapped onto parameters of the links connecting the corresponding qubits. However, this process is subject to a variety of constraints in practice, which limit the applicability of the quantum annealer. First, the number of available hardware qubits is relatively small, which restricts the admissible problem size of eq.~\eqref{eq:hamiltonian}. Second, the connectivity of the hardware qubits on the D-Wave quantum chip is limited and local. Therefore, for two logical qubits $x_i$ and $x_j$ with $a_{ij} \neq 0$ in eq.~\eqref{eq:hamiltonian}, it is not guaranteed that a physical link between qubits $i$ and $j$ exists on the quantum hardware. To alleviate this issue, a \textit{minor embedding} of the connectivity structure of eq.~\eqref{eq:hamiltonian} onto the D-Wave qubit hardware graph is needed, where one logical qubit is represented by a connected set of physical qubits, called a \textit{chain}. However, the existence of chains often results in a severe reduction in the number of available qubits \cite{Choi2008, Chapuis2017}. Third, excessive chain lengths may cause the solution quality to decrease \cite{Marshall2020, Grant2021, Cai2014}. Therefore, the quality of the solutions found by the D-Wave annealer is typically higher for smaller problems and, specifically, for problems with shorter chain lengths.

This research shows that, by combining an exact classical graph decomposition algorithm  \cite{Pelofske2019mc} with a method for solving multiple smaller problems on the quantum annealing in parallel \cite{Pelofske2022parallel}, one can solve accurately problems of sizes larger than what fits on the quantum annealer hardware. We focus on solving the Maximum Clique (MC) problem, an NP-hard problem with important applications in network analysis, bioinformatics, and computational chemistry \cite{Pelofske2019mc}. We test the method on Erd\H{o}s–R\'{e}nyi random graphs \cite{ErdosRenyi1960} of up to $120$ vertices. The decomposition algorithm splits a given graph instance for which MC is to be computed into subproblems that are strictly smaller than the input graph. By applying the decomposition in a recursive fashion, an arbitrary sized input can be broken down to subproblems suitable for the D-Wave quantum annealer. Importantly, though the algorithm of \cite{Pelofske2019mc} has an exponential worst-case complexity, it is  exact, in the sense that an optimal solution of the input problem is guaranteed given all generated subproblems are solved exactly by the quantum annealer.

This work is an extension of the two articles of \cite{Pelofske2019mc} and \cite{Pelofske2021decomposition}, in which the decomposition algorithm for the MC problem was introduced. However, both previous articles only investigated the decomposition from a classical point of view, in the sense that the generated subproblems were never actually solved on the D-Wave quantum annealer. In contrast to the previous works, in this contribution we employ disjoint clique embeddings to actually solve the generated problems simultaneously on the D-Wave quantum hardware. This is possible due to the advancement in the quantum annealing hardware, and particularly the availability of the D-Wave Advantage computers with over 5000 qubits, and the parallel quantum annealing method introduced in \cite{Pelofske2022parallel}. Parallel quantum annealing can be equivalently referred to as \emph{tiling}\footnote{\url{https://dwave-systemdocs.readthedocs.io/en/samplers/reference/composites/tiling.html}}. We note that a similar paradigm of parallel computation exists for universal gate model quantum computers \cite{9749894, Niu2023enablingmulti, 10.1145/3352460.3358287}. 
We investigate the performance and solution quality of the proposed quantum algorithms to solve the MC problem on multiple graphs. Performance measures include the number of subgraphs generated as a function of the recursion cutoff, the number of disjoint clique embeddings used for parallel quantum annealing, the success rate of finding ground states (optimal solutions) at certain subgraph sizes, and time-to-solution (TTS) measurements for finding a maximum clique.

This work is structured as follows. After a literature review in Section~\ref{sec:literature}, Section~\ref{sec:methods} introduces the MC problem and its formulation as a QUBO and introduces the  decomposition method for MC, as well as discusses a generalization of the TTS metric for solving MC instances with parallel embeddings. All experimental results on using D-Wave in connection with decomposition and parallel embeddings can be found in Section~\ref{sec:experiments}. The article concludes with a discussion in Section~\ref{sec:discussion}.

\subsection{Literature review}
\label{sec:literature}
Exact algorithms for solving NP-hard problems on real-life problems of practical importance have received continuous attention in the literature~\cite{dimacs1996, dimacs2000, Woeginger2008}. One can broadly distinguish between three flavors of such methods: exact exponential-time algorithms to decompose NP-hard problems, (polynomial-time) algorithms for special cases of NP instances, and hybrid algorithms relying on both decomposition and quantum annealing.

The idea of an exact decomposition for NP-hard problems, which also lays at the heart of the present contribution, dates back to at least \cite{Tarjan1985}. Similar techniques for other NP-hard problems such as graph coloring are known \cite{Rao2008}. For MC, the exact decomposition algorithm that serves as the basis of the present work has been proposed in \cite{Chapuis2017}. In \cite{Pelofske2019mc}, the authors consider several techniques to speed up the decomposition by pruning subproblems that cannot contribute to the optimal solution, for instance by computing bounds on clique sizes in subproblems. Although bounding and pruning subproblems can considerably speed up the computations and allow one to solve problems of practical importance exactly, there is no guarantee that exact decomposition techniques terminate in reasonable time, and they do not asymptotically lower the exponential runtime complexity.

For the Maximum Independent Set problem, the complimentary problem of MC, a variety of exact algorithms are available \cite{Robson1986, Robson2001, Fomin2006, Xiao2013}. Additionally, special cases can be solved in polynomial-time \cite{Minty1980, Grotschel1988, Dabrowski2011}. Some known algorithms for such special cases rely on graph decomposition \cite{Giakoumakis1997, Courcelle2000}. Similarly to before, the aforementioned exact techniques still have an exponential runtime complexity, and the special cases that can be solved in polynomial time are rather specific, meaning they usually do not apply to problems of practical importance.

The algorithm of \cite{Pelofske2019mc} has been generalized in \cite{Pelofske2019minvc} to decompose large Minimum Vertex Cover (MVC) problems, a related problem to MC, and turned into a framework for decomposing certain NP-hard graph problems that does not explicitly consider quantum annealing \cite{Pelofske2021decomposition}. Further exponential-time algorithms for MVC have been proposed \cite{Balasubramanian1998, StegeFellows1999, Chen2000, Chen2001}, including some aiming to reduce the size of an MVC instance \cite{Niedermeier20070UB, Chen2010}, and some considering the weighted MVC problem \cite{NiedermeierRossmanith2003, Xu2016}.

The related problem of enumerating all maximum cliques of a graph has been addressed in the literature with the algorithms of \cite{BronKerbosch1973} and \cite{CarraghanPardalos1990}, which both do not rely on graph decomposition. A parallel version of the algorithm of \cite{BronKerbosch1973} can be found in \cite{Rossi2015}.

The present work falls under the area of branch-and-prune algorithms, specifically those that employ a different solver once the generated subproblems of the decomposition are sufficiently small. A survey of such algorithms can be found in \cite{Hou2014} and \cite{Morrison2016}. Recent works that specifically employ the D-Wave quantum annealer in connection with a classical decomposition include \cite{Bass2020}, who use the heuristic D-Wave tool ``QBsolv''.

Although having their merits for practical use, the aforementioned algorithms cannot solve arbitrary instances in time that is asymptotically lower than exponential. Moreover, techniques using decomposition in connection with quantum annealing do not give a guarantee of optimality, as solutions of subproblems solved via quantum annealing are of a heuristic nature.

Finally, a survey of classical simplification techniques for QUBO problems, including problem-agnostic decomposition algorithms for QUBOs that are applicable in special cases, can be found in \cite{Boros2002} and \cite{Boros2006}.

\section{Methods}
\label{sec:methods}
This section introduces the MC problem that we aim to solve (Section~\ref{sec:methods_MC}) and the DBK algorithm, which is able to decompose an MC instance recursively into smaller instances of prespecified size (Section~\ref{sec:methods_DBK}). Moreover, we introduce the D-Wave Advantage System~4.1 and its hardware (Section~\ref{sec:methods_DWave}), as well as the idea of parallel quantum annealing (Section~\ref{sec:methods_parallel_QA}), which allows one to solve several problem instances in a single call to the quantum annealer. A generalization of the time-to-solution (TTS) metric for measuring the performance of the DBK algorithm is introduced in Section~\ref{sec:TTS}. 

\subsection{The Maximum Clique problem}
\label{sec:methods_MC}
Let $G=(V,E)$ be an undirected graph with vertex set $V$ and edge set $E \subseteq V \times V$. A subgraph $G(S)$ of $G$ induced by a subset $S \subseteq V$ is called a \textit{clique} of $G$ if it is a complete subgraph, that is, if $(v,w) \in E$ for all $v,w \in S$, $v \neq w$. A \textit{maximum clique} of $G$ is a clique of $G$ of maximum size.

Many NP-hard problems can be easily reformulated as QUBO or Ising problems of the form of eq.~\eqref{eq:hamiltonian}. This includes, for instance, graph coloring, graph partitioning, or the MC problem \cite{Lucas2014}. As given in \cite{Pelofske2019mc}, the QUBO formulation for the MC problem on a graph $G=(V,E)$ is given by
\begin{align}
    H_{MC} = -A\sum_{v \in V} x_v + B\sum_{(u,v) \in \overline{E}} x_u x_v,
    \label{eq:MC}
\end{align}
where $\overline{E}$ denotes the edge set of the complement graph of $G$ and the constants $A>0$ and $B>0$ need to satisfy $A<B$. Without loss of generality, we fix $A=1$ and $B=2$ in the remainder of this work. Each binary variable $x_v \in \{0,1\}$ for $v \in V$ indicates if vertex $v$ belongs to the maximum clique ($x_v=1$) or not ($x_v=0$).

\subsection{The DBK algorithm}
\label{sec:methods_DBK}
The DBK algorithm \cite{Pelofske2019mc, Pelofske2021decomposition, Pelofske2019minvc} was designed to decompose an input graph recursively into smaller subgraphs such that, in each recursion level, (a) each generated subgraph is strictly smaller than the graph that was split, and (b) the maximum clique of the input graph can be reconstructed exactly from the maximum cliques determined for the leaf graphs, given all subgraphs can be solved exactly. The name of the algorithm stands for decomposition, bounds, and k-core. These three components work as follows.

\begin{figure}[t]
    \centering
    \includegraphics[width=0.5\textwidth]{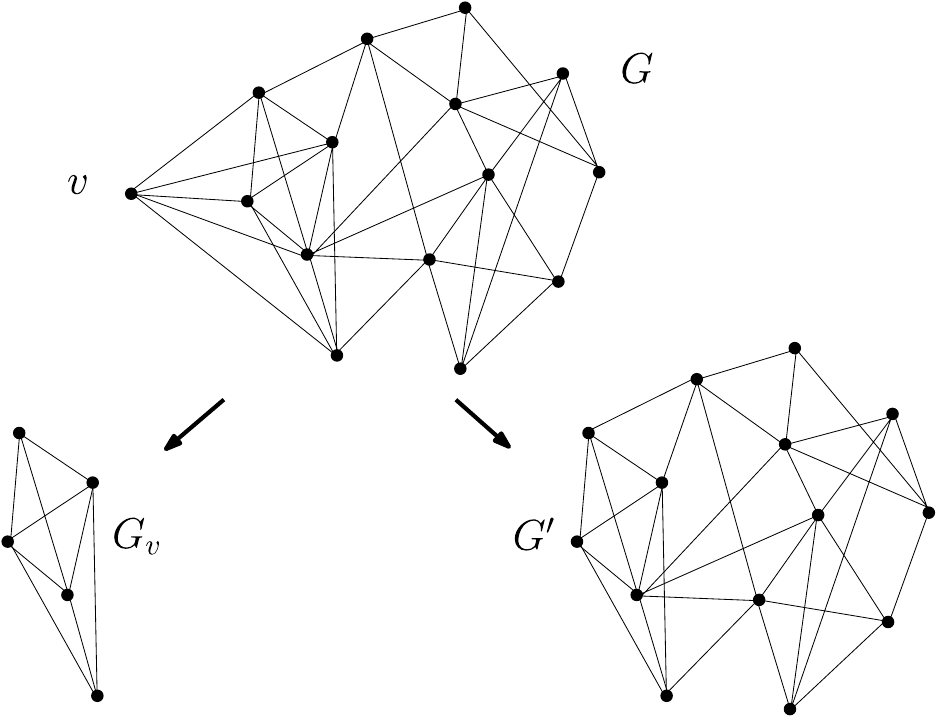}
    \caption{Illustration of the DBK vertex splitting applied to vertex $v$, resulting in the induced subgraph $G_v$ of $v$, and a graph $G'=(V',E')$ with $V'=V \setminus \{v\}$ and all edges incident to $v$ removed from $E$. Figure taken from \cite{Pelofske2019mc}.\label{fig:dbk}}
\end{figure}

First, the decomposition of an input graph $G$ is illustrated in Figure~\ref{fig:dbk}. We choose a random vertex $v \in G$ and use it to split $G$ into two graphs, the subgraph $G_v$ induced by the neighbor vertices of $v$, and the subgraph $G'$ obtained by removing $v$ from $G$ including all its incident edges. If $v$ is part of a maximum clique then that maximum clique, minus vertex $v$, will be contained in $G_v$. Hence, the size of the maximum clique of $G_v$ will be one less than the size of the maximum clique of $G$. On the other hand, if $v$ is not part of a maximum clique, then $v$ can be removed from $G$, resulting in a new graph $G'$ that contains the same maximum clique as $G$. Therefore, the maximum clique can be reconstructed after determining the maximum clique in both $G_v$ and $G'$, with each of these graphs being strictly smaller than $G$. The decomposition is valid irrespective of the choice of the vertex $v$ used for splitting the graph, though some choices might be advantageous in that they make the algorithm terminate faster. As shown in \cite{Pelofske2019mc, Pelofske2021decomposition}, choosing $v$ as the vertex of lowest degree yields the fastest solutions.

Second, before recursing into any of the generated subgraphs, the size of the clique contained in a subgraph can be lower and upper bounded using various techniques. For instance, the chromatic number computed with a fast heuristic (we use the Python module \textit{NetworkX} of \cite{Networkx}), and the upper bound of \cite{Budinich2003} provide efficiently computable upper bounds. Likewise, a trivial lower bound on any subproblem in the decomposition tree is obtained by simply solving the other subbranch. These upper and lower bounds are employed in the DBK algorithm. These bounds were selected due to their low runtime complexity and good performance compared to other bounds (thus allowing one to prune many subproblems).

Third, the DBK algorithm attempts to simplify any newly generated subproblem in the decomposition tree before solving it further in a recursive fashion. After exploring several techniques, the \textit{k-core} algorithm was selected in \cite{Pelofske2019mc}. This algorithm works as follows. The $k$-code of a graph is the subgraph consisting of all vertices having degree at least $k$. Clearly, at any point during the decomposition, one can simplify a subgraph by replacing it with its $k$-core, where $k$ is set to the largest clique number found so far, thus potentially reducing the size of some of the subproblems before continuing the recursion.

The complete algorithm is summarized in Algorithm~\ref{algo:DBK}, which uses a stack called \textit{subgraphs} instead of a recursion. Starting from the input graph $G$, Algorithm~\ref{algo:DBK} first queries a lower bound on the maximum clique and then simplifies $G$ by replacing it with its $k$-core. Next, $G$ is split at the lowest degree vertex of $G$. Both resulting subgraphs, denoted $g'$ and $g''$, are then examined recursively. After replacing them with their $k$-core as done before, the maximum clique size is lower bounded using the MCs found in $g'$ and $g''$ and $k$ is updated if necessary. If the size of a subgraph is at most $L$, the size cutoff at which we attempt to solve the maximum clique problem, we query the provided solver method, otherwise a subgraph is added to the stack \textit{subgraph} for further decomposition. Though proven to be exact, the DBK algorithm has a worst-case exponential runtime (which is to be expected since the MC problem is NP-hard).

The DBK algorithm can be specifically defined based on what Maximum Clique solver method is used. For example a heuristic Maximum Clique solver could be used, in which case, despite the decomposition being exact, the solution may not be optimal. In this case we use two different versions of DBK. 

The first we call \emph{DBK-pQA}; here the Maximum Clique problem is sampled using the D-Wave Advantage System 4.1 quantum annealer with parallel quantum annealing. In this case the solver method that is used in the DBK algorithm (line 16 of Algorithm~\ref{algo:DBK}) is querying the quantum annealer for $1,000$ samples, which are then post processed into solutions for the maximum clique problem. The largest clique found in those $1,000$ samples is returned to the main DBK algorithm. The \emph{DBK-pQA} method is a hybrid quantum-classical algorithm. 

The second variant of DBK we call \emph{DBK-fmc}. In this case the Maximum Clique solver method (line 16 of Algorithm~\ref{algo:DBK}) is the \textit{Fast Maximum Clique} solver of \cite{fmc}, which is an exact classical solver. Therefore the \emph{DBK-fmc} algorithm is guaranteed to find the Maximum Clique of the given graph. 

\begin{algorithm}[t]
    \caption{DBK-Generic}
    \label{algo:DBK}
	\begin{algorithmic}[1]
        \Require graph $G=(V,E)$, recursion cutoff $L$
        \Ensure clique size $k$
        \State $k\leftarrow$ lower bound on maximum clique of $G$
        \State $G \leftarrow$ $k$-core of $G$
        \State subgraphs $\leftarrow$ $\{G\}$
        \While {subgraphs $\neq \emptyset$}
            \State $g$ $\leftarrow$ subgraphs.pop()
            \State $v$ $\leftarrow$ lowest degree vertex of $g$
            \State Partition $g$ at vertex $v$ into $g'$ and $g''$
            \For {$g^* \in \{g',g''\}$}
                \State $g^* \leftarrow$ $k$-core of $g^*$
                \If {lower bound on maximum clique of $g^*$ exceeds $k$}
                    \State update $k$
                \EndIf
                \If {size of $g^*$ exceeds $L$} 
                    \State subgraphs.push($g^*$) 
                \ElsIf {upper bound on maximum clique of $g^*$ exceeds $k$} 
                    \State Call solver method on $g^*$
                    \State Update $k$ while taking into account previously removed vertices
                \EndIf
            \EndFor
    	\EndWhile
    	\State \textbf{return} $k$
    \end{algorithmic}
\end{algorithm}

\subsection{The D-Wave Advantage System~4.1}
\label{sec:methods_DWave}
The hardware connectivity of the D-Wave Advantage System systems is referred to as Pegasus \cite{Boothby2020}, and the connectivity graphs of the D-Wave 2000Q systems are referred to as Chimera \cite{boothby2016fast, vert2019limitations}. Both of these connectivity graphs are relatively sparse, and they do not allow for the direct mapping of problem QUBO or Ising formulations with arbitrary connectivity onto the hardware. As an alternative to the direct embedding, a \emph{minor embedding} allows a problem with arbitrary connectivity (up to the hard limit of the size of the hardware) to be embedded \cite{Choi2008}. In a minor embedding, a representation of the logical problem is created in which sets of physical qubits are linked together into \emph{chains} \cite{king2014algorithm}, where each chain represents a single logical qubit. Usually, the programmed weight of the logical variable is uniformly distributed across the chain qubits, although alternative weight distributions are possible \cite{https://doi.org/10.48550/arxiv.2202.03044, barbosa2020optimizing}. Computing a minor embedding with minimum chain length is NP-hard, but there are heuristics that can be used in order to efficiently generate minor embeddings \cite{Cai2014}. Another viable method to circumvent repeatedly generating minor-embeddings is to generate a fixed clique minor-embedding \cite{boothby2016fast}, which then allows one to embed an arbitrary graph of any connectivity so long as its size is at most the one of the clique embedding.

One of the disadvantages of minor embeddings is that at large chain lengths, the chains may begin to disagree on the logical variable state when the final state of the qubits is read out. We call these instances \emph{broken chains}. Broken chains typically indicate poor solution quality \cite{grant2022benchmarking, Marshall_2022}, and the solutions with broken chains also need to either be repaired in some way or discarded entirely. Repairing broken chains is referred to as \emph{unembedding}  \cite{unembedding-ICRC}. A simple method for resolving broken chains, and thus to form a logical variable state, is to simply take the \emph{majority vote} on the qubit states of all physical qubits in a chain. Note that in the edge case of a broken chain being evenly split between states then a random choice with $p=0.5$ is used to resolve the chain. In this research we always apply the majority vote unembedding method. 

\subsection{Parallel Quantum Annealing}
\label{sec:methods_parallel_QA}
One of the other disadvantages of minor embeddings is that a fixed clique embedding does not typically make maximal use of the hardware available (i.e., many qubits and couplers available on the hardware stay idle as nothing is mapped onto them). This problem, in conjunction with poorer solution quality at larger embedding sizes, gives rise to the natural idea of embedding multiple smaller (disjoint) cliques onto the quantum annealing hardware and thus solving multiple minor-embedded problems on the quantum annealer during the same anneal \cite{Pelofske2022parallel}. This idea is referred to as \emph{parallel quantum annealing}, which makes better use of the available hardware compared to the sequentially solving of all individual problems. Parallel quantum annealing can be applied in order to solve a set of different QUBOs on a quantum annealer, or it can be used to solve the same QUBO multiple times on the quantum annealer. In this research we apply the method of solving the same maximum clique QUBO on as many embeddings (this is determined by the size of the QUBOs) as can be fit using the heuristic minor-embedding tool \emph{minorminer} \cite{Cai2014}.

\begin{figure}[h]
    \centering
    \includegraphics[width=0.25\textwidth]{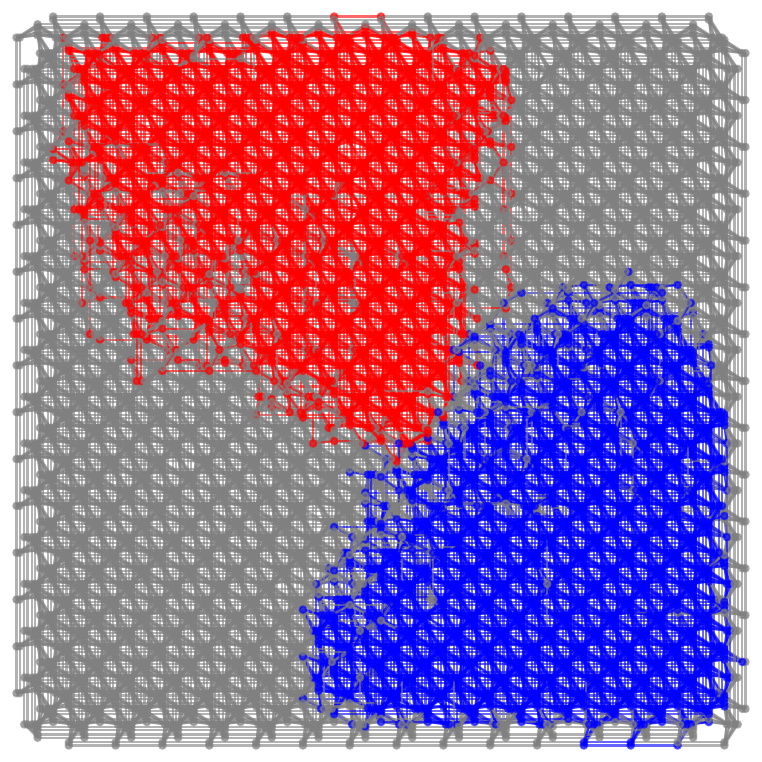}
    \includegraphics[width=0.25\textwidth]{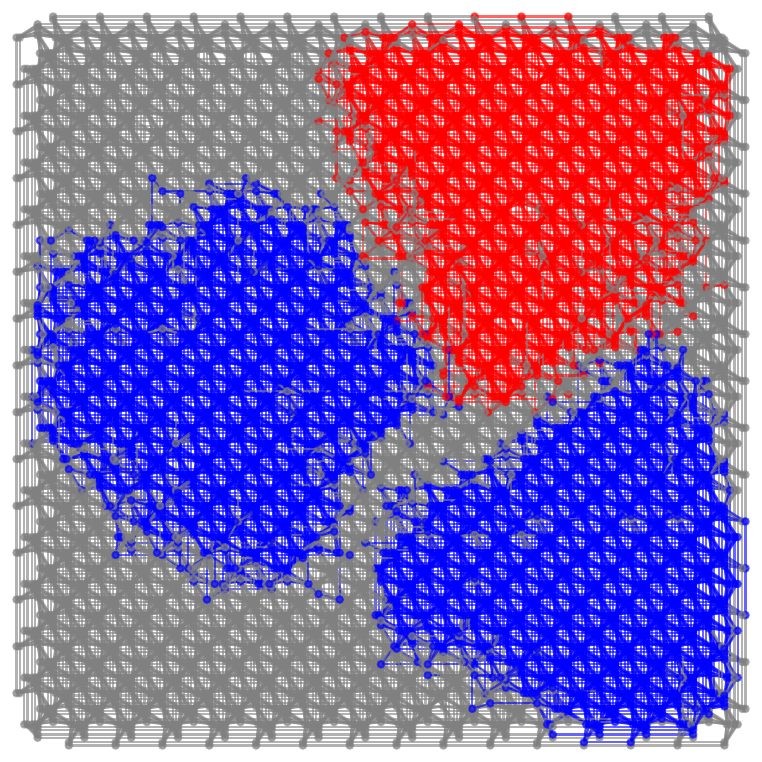}
    \includegraphics[width=0.25\textwidth]{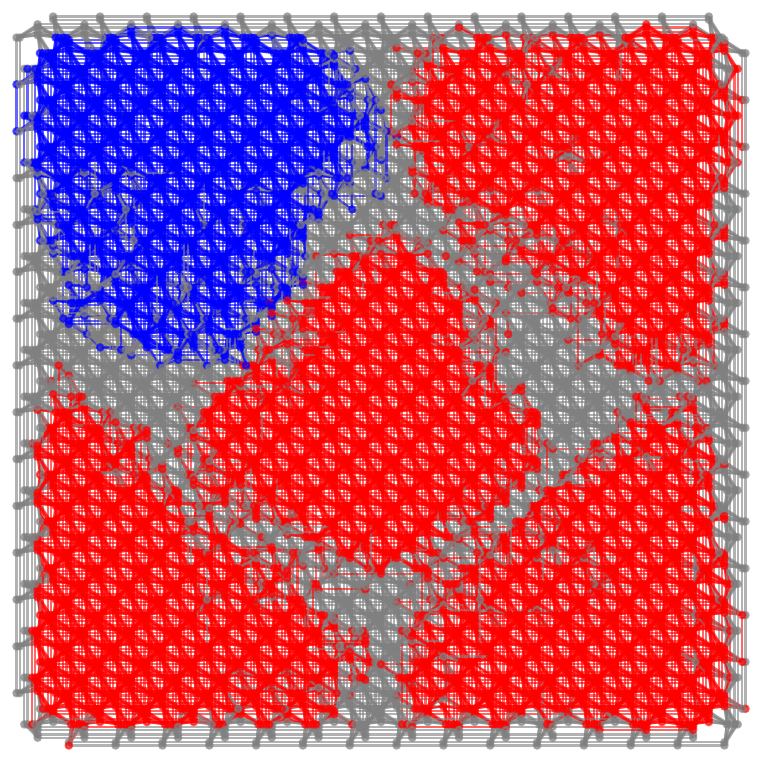}\\
    \includegraphics[width=0.25\textwidth]{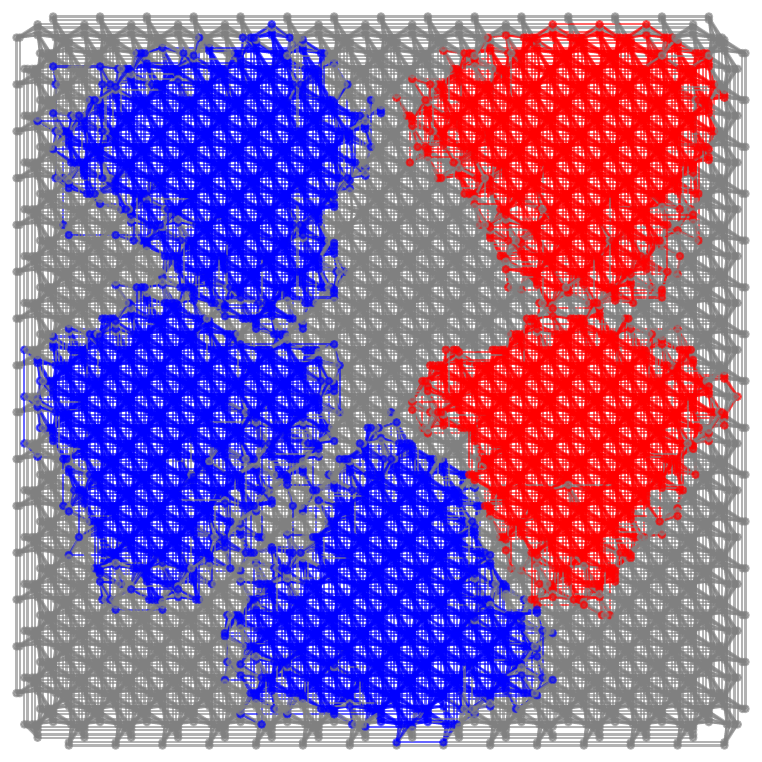}
    \includegraphics[width=0.25\textwidth]{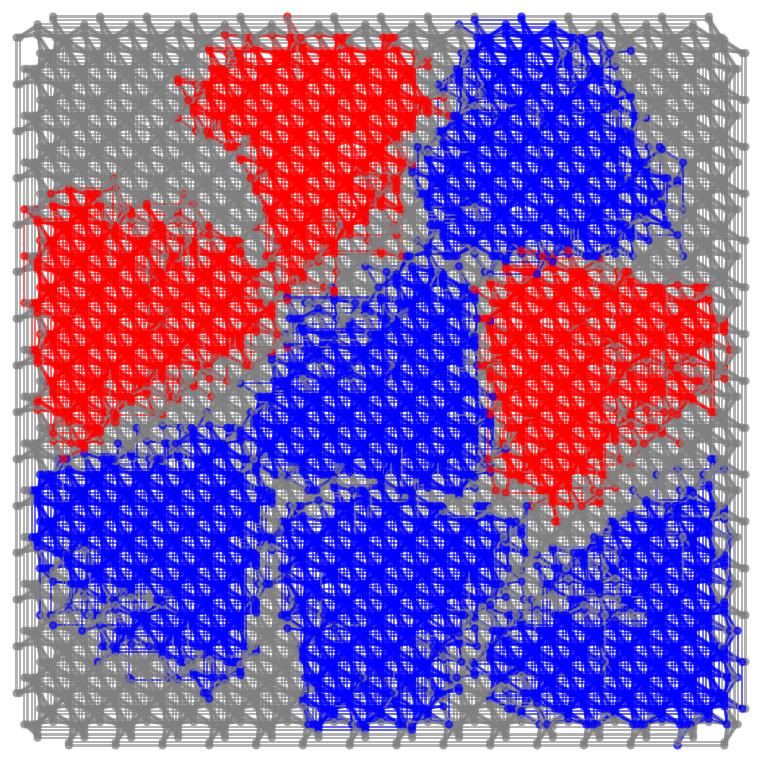}
    \includegraphics[width=0.25\textwidth]{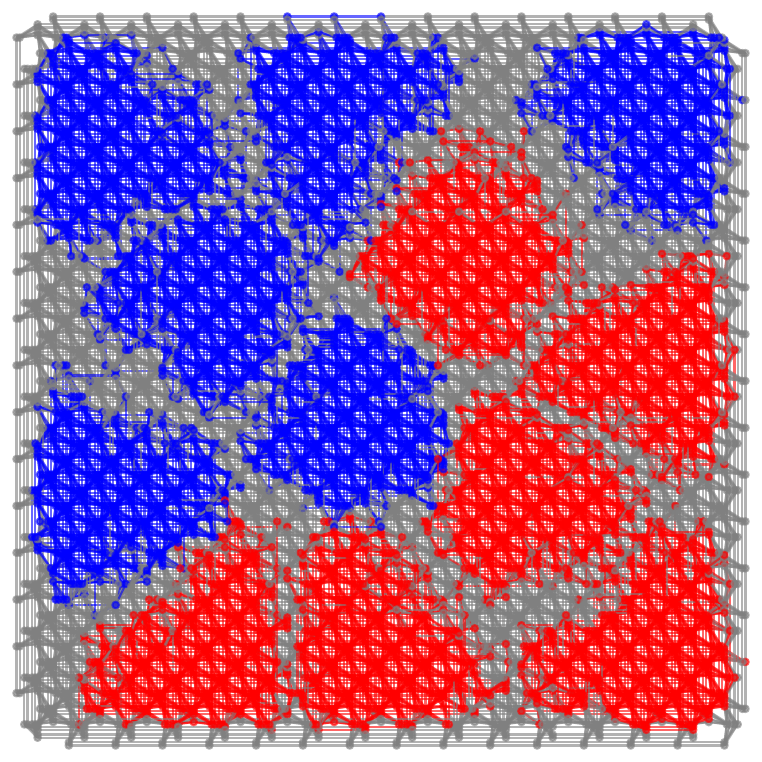}
    \caption{Disjoint minor embeddings for parallel problem solving on the quantum annealer. The chip topology is for the D-Wave Advantage System 4.1. The minor-embeddings are of cliques of sizes $N \in \{100, 90, 80, 70, 60, 50\}$ from top left to bottom. Red and Blue coloring is used (randomly) in order to help to visually differentiate neighboring minor embeddings. \label{fig:parallel_embeddings}}
\end{figure}

The quantum annealing backend used is the D-Wave Advantage System~4.1, hereafter referred to as D-Wave. Figure~\ref{fig:parallel_embeddings} shows the disjoint clique minor embeddings on the Pegasus connectivity graph of sizes ranging from two $100$-vertex clique embeddings to twelve $50$-vertex clique embeddings. Each of the disjoint embeddings can be used to solve either the same QUBO repeatedly in the same anneal, or to solve different QUBOs.

For usage in conjunction with the DBK algorithm, we solve each subproblem that DBK generates by computing the QUBO of the MC problem for that graph and embedding the QUBO as many times as possible given the fixed disjoint clique embeddings displayed in Figure~\ref{fig:parallel_embeddings}. The size of the clique embeddings being used is always equal to the DBK cutoff value used. This increases the probability that a maximum clique will be found during a single D-Wave call \cite{Pelofske2022parallel}. This is different from previous DBK approaches \cite{Pelofske2019mc, Pelofske2021decomposition} which solved subproblems using sequential quantum annealing.

While running the experiments in the context of the DBK algorithm, there is a potential inefficiency in the embedding usage, which is due to the fact that the sizes of the subgraphs generated by the DBK algorithm can be less than the DBK cutoff value. In such a case, the utilized embedding is larger than required, thus resulting in inefficient usage of the available qubits and unnecessarily long chains, which in turn potentially decreases the solution quality. However, for the purposes of being able to solve multiple problems simultaneously, this is a reasonable cost to take on.

\subsection{The time-to-solution (TTS) metrics}
\label{sec:TTS}
Since the D-Wave quantum annealer is a probabilistic solver and because of its imperfect hardware, it only samples a ground state solution for a QUBO or Ising problem with a certain probability. Therefore, many samples are usually necessary to obtain an optimal, or at least a sufficiently good, solution. In order to estimate the time it takes for D-Wave to find an optimal solution, one can use the so-called time-to-solution (TTS) metric \cite{tts1,tts2}. TTS is a measure of the time it takes to reach an optimal solution with a $99$ percent probability.

When using parallel quantum computing, i.e., when solving $N \in \N$ problem instances simultaneously on the D-Wave hardware, the standard method to compute the TTS does not apply. 
Let $A_i \in \N$ be the number of samples for problem $i \in \{1,\ldots,N\}$, let $T_{\text{QPU}_i}$ be the QPU access time (in seconds), and $T_{\text{unembed}_i}$ be the classical processing time (in seconds) to unembed the solution vectors across all disjoint minor-embeddings. Since the same problem is embedded multiple times on the hardware chip, it is possible that a single anneal might have found the ground state solution multiple times. However, for the purposes of computing the ground state probability $p_i$, we solely count the number of samples (among the total $A_i$ samples) that found the ground state solution at least once. If $p_i = 0$ we can not compute $\text{TTS}_\text{opt}$. The quantity $p_i$ is also called the ground state probability (GSP) for subgraph $i \in \{1,\ldots,N\}$. Lastly, we record the DBK processing time required to carry out the DBK decomposition (i.e., excluding the time required to solve each subgraph), denoted as $T_\text{DBK\_proc\_time}$. 
We obtain the formula
\begin{align}
    \text{TTS}_\text{opt} = T_\text{DBK\_proc\_time} + \sum_{i=0}^N \frac{1}{A_i} \cdot \left( T_{\text{QPU}_i} + T_{\text{unembed}_i} \right) \cdot \frac{\log(0.01)}{\log(1-p_i)},
    \label{eq:TTS}
\end{align}
if $0<p_i<1$. The coefficient $\frac{\log(0.01)}{\log(1-p_i)}$ corresponds to the number of times the algorithm should be run for the corresponding subgraph to ensure $99\%$ probability of finding an optimal solution. In the special case where $p_i = 1$ for some $i \in \{1,\ldots,N\}$, we set the TTS for that subgraph to be $(T_{\text{QPU}_i} + T_{\text{unembed}_i})/A_i$.
Also, since we use a fixed minor-embedding, we can set the minor embedding computation time to zero because the embedding can be pre-computed.

Measuring the time-to-solution of a probabilistic algorithm is a useful way of comparing it to other algorithms. However, in computing the TTS for an algorithm that uses a quantum annealer, there are some important assumptions to note. First, when estimating the time to run an anneal (and the time to unembed that solution), we are assuming that the QPU time (and unembedding time) are proportional to the number of samples. This assumption is not necessarily true, in particular because one can gain efficiency by batching samples together into the same job. Second, the TTS formula aims to compute the minimum amount of time that is required to reach the optimal solution with $99\%$ confidence. However this does not take into account that determining the optimal number of samples to take in order to reach this ideal TTS in general is not known before actually running the algorithm. Therefore, quantifying TTS in this manner gives a lower bound on the \emph{ideal} computation time.

When using a quantum annealer in order to solve optimization problems in practice, it is easier to set the number of samples per problem to some fixed number. However, this means that when  the quantum annealer is used as a subsolver in a larger hybrid algorithm (for instance DBK), then the optimal TTS given by formula \eqref{eq:TTS} will not be achieved. In this case, we can consider computing the TTS of the DBK algorithm, where the success rate $p$ is the number of trials where DBK succeeded in finding the maximum clique and $T_{\text{QPU}_i}$ and $T_{\text{unembed}_i}$ will be the same for all problems (since they only depend on the number of samples $A_i$). Importantly in this case the number of samples to use when querying the quantum annealer $A_i=A$ can be selected by the user. The resulting TTS formula for fixed number of samples is
\begin{align}
    \text{TTS}_\text{fixed} = \left( T_\text{QPU}+T_\text{classical} \right) \cdot \frac{\log(0.01)}{\log(1-p)},
    \label{eq:TTS_real_time_DBK}
\end{align}
Here, $T_\text{classical}$ is the classical time used by D-Wave, e.g., mapping the problem onto the D-Wave hardware and postprocessing the samples.

In the case $p=1$, we set $\text{TTS}_\text{fixed} = T_\text{QPU}+T_\text{classical}$. If $p=0$, then $\text{TTS}_\text{fixed}$ is not defined.

\section{Experiments}
\label{sec:experiments}
This section presents some experimental results. After introducing the experimental setup in Section~\ref{sec:setup}, we apply the DBK-fmc algorithm of Section~\ref{sec:methods_DBK} to a set of test graphs. We study how the number of subgraphs as well as their size and density behaves during the decomposition (Section~\ref{sec:subgraphs}), and we investigate the probability of finding an optimal solution at the leaf level of the decomposition tree, when using parallel quantum annealing, as a function of the DBK-fmc cutoff value (Section~\ref{sec:groundstates}). Moreover, we consider a comparison of the DBK-fmc algorithm where the sub-problems are solved using parallel quantum annealing and the classical FMC solver using a modified TTS measure that emulates an optimal usage of quantum annealing in conjunction with DBK (Section~\ref{sec:comparison}). The experiments conclude with an application of the DBK-pQA algorithm where a quantum annealer is used as a real time subsolver with a fixed number of samples (Section~\ref{sec:experiments_real_time_DWave_solver}).

\subsection{Experimental set-up}
\label{sec:setup}
All experiments are performed on a series of $60$ Erd\H{o}s–R\'{e}nyi random graphs \cite{ErdosRenyi1960}. Each of these $60$ graphs has $120$ vertices and density sampled from a continuous uniform distribution in $[0.1,0.9]$. Additionally, we ensure that each of the test graphs is connected, meaning there is a path between any pair of vertices.

We apply the DBK algorithm to each of the $60$ graphs using different cutoff sizes for the subgraphs. To be precise, we study the behavior of the DBK algorithm for the cutoff values $\{ 110, 100, 90, 80, 70, 60, 50\}$, where the solver used is the classical \textit{Fast Maximum Clique} solver of \cite{fmc}. Then, we investigate under what parameters the D-Wave quantum annealer would be able to solve the generated subgraph problems (as well as what the precise Time-to-Solution characteristics are). In this way, similarly to \cite{Pelofske2021decomposition, Pelofske2019minvc, Pelofske2019mc}, we can emulate the expected quantum annealer behavior in the DBK algorithm. Next, using the information gained from the the previous step we run DBK using the quantum annealer as the real-time subsolver.

The D-Wave settings we use are \emph{annealing time} $50$ microseconds, $1000$ samples per backend call, \emph{readout thermalization} $0$ microseconds, \emph{programming thermalization} $0$ microseconds, and the \emph{reduce intersample correlation} Boolean flag is set to True. The chain strength value is dynamically computed based on the individual QUBO properties using the D-Wave Ocean SDK function \textit{uniform torque compensation} \cite{uniform-torque-compensation} (with a user-specified prefactor of $0.2$). The uniform torque compensation method attempts to provide a chain strength that reduces the number of broken chains, chosen as the square root of the mean of the quadratic couplers of the QUBO. We found empirically that for the large Maximum Clique minor embedded problems, a uniform torque compensation prefactor less than $0.5$ and greater than $0.1$ gave the overall best results; we chose to use $0.2$, which resulted in favorable approximation ratios overall (see Figure \ref{fig:approximation_ratios}). Because the settings across these experiments are constant, the resulting QPU time (specifically \emph{qpu-access-time}) is relatively constant in these experiments at about $1$ second per backend call. 

In Figures \ref{fig:DBK_decomposition_summary}, \ref{fig:failure_rate}, and \ref{fig:TTS} we color code lower input graph densities with dark blue, intermediate densities ($0.5$) are colored orange and yellow corresponds to higher input graph densities (before decomposition).

\subsection{Number, density, and size of subgraphs with DBK-fmc}
\label{sec:subgraphs}
\begin{figure}[h]
    \centering
    \includegraphics[width=0.49\textwidth]{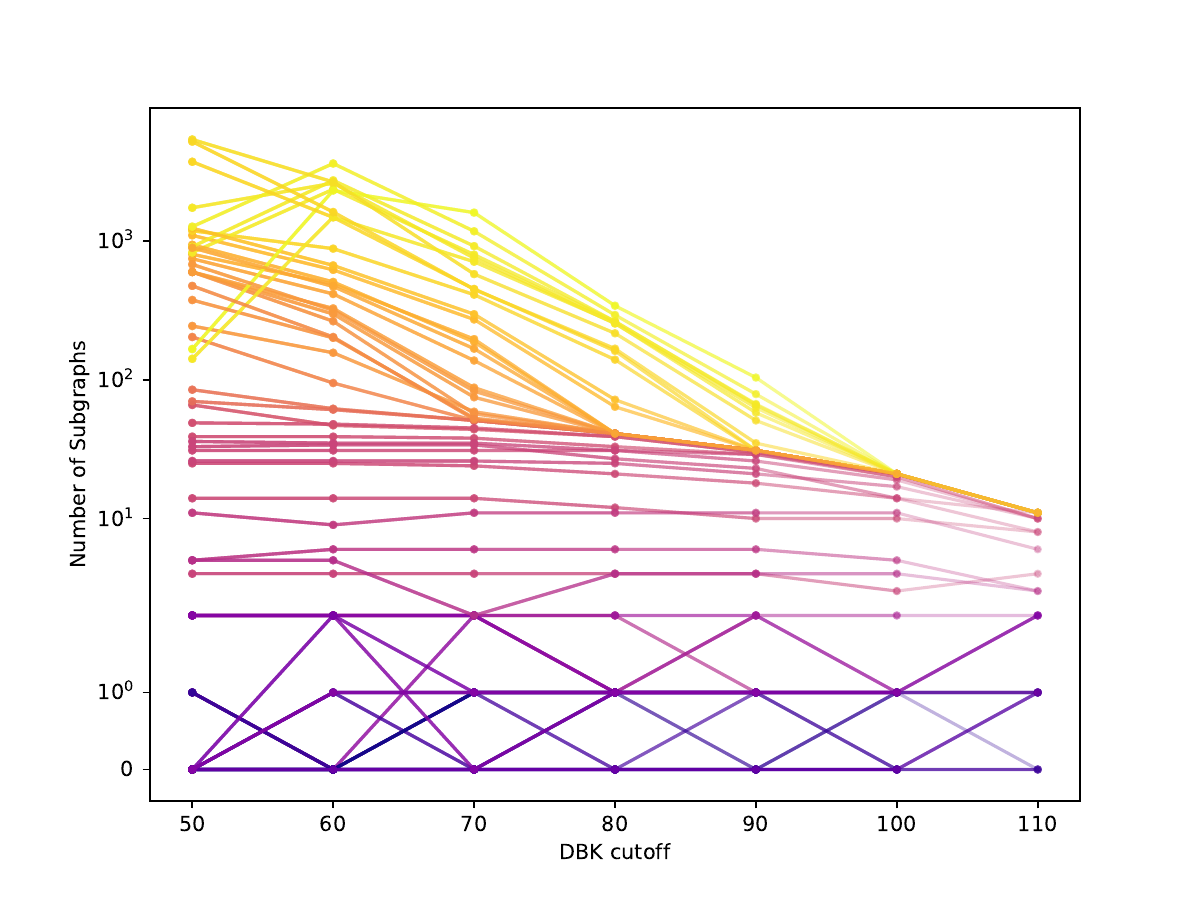}
    \includegraphics[width=0.49\textwidth]{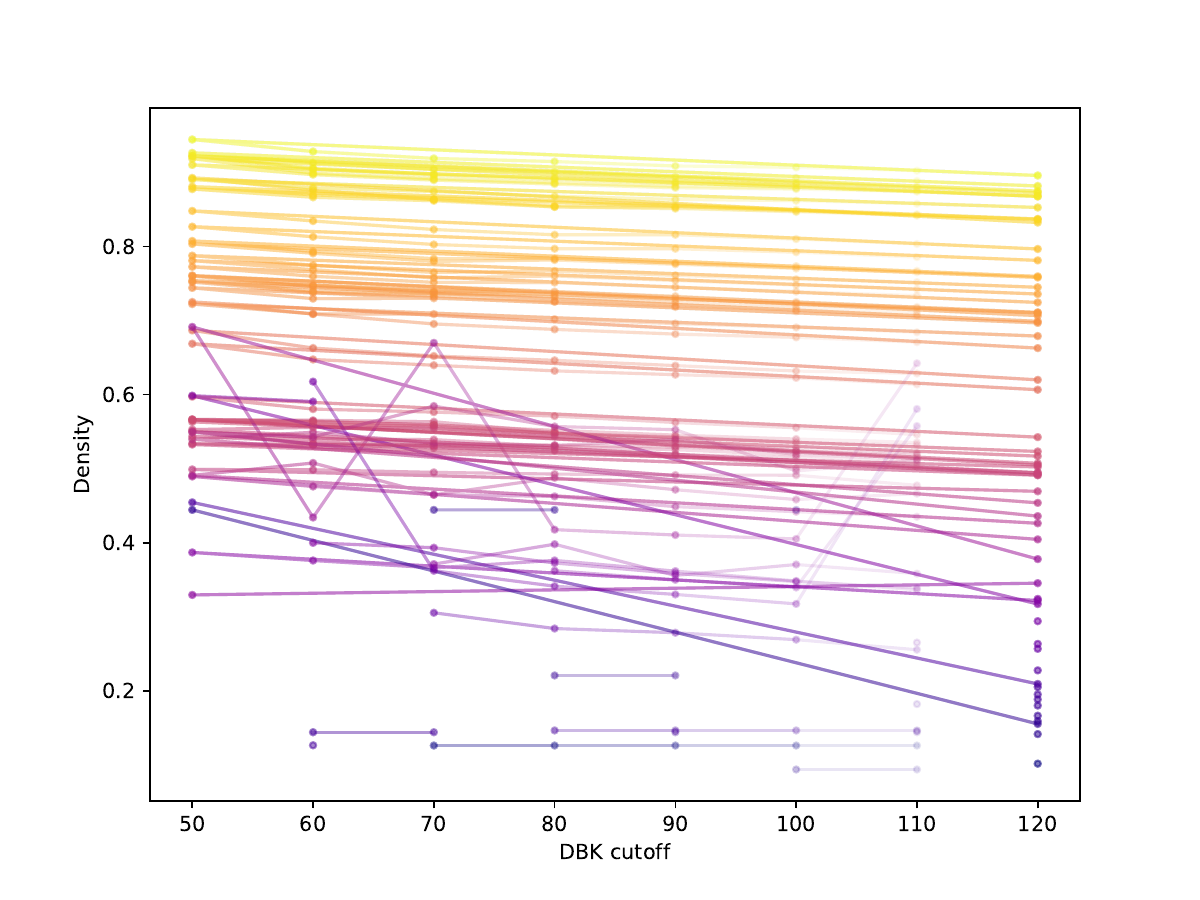}
    \includegraphics[width=0.49\textwidth]{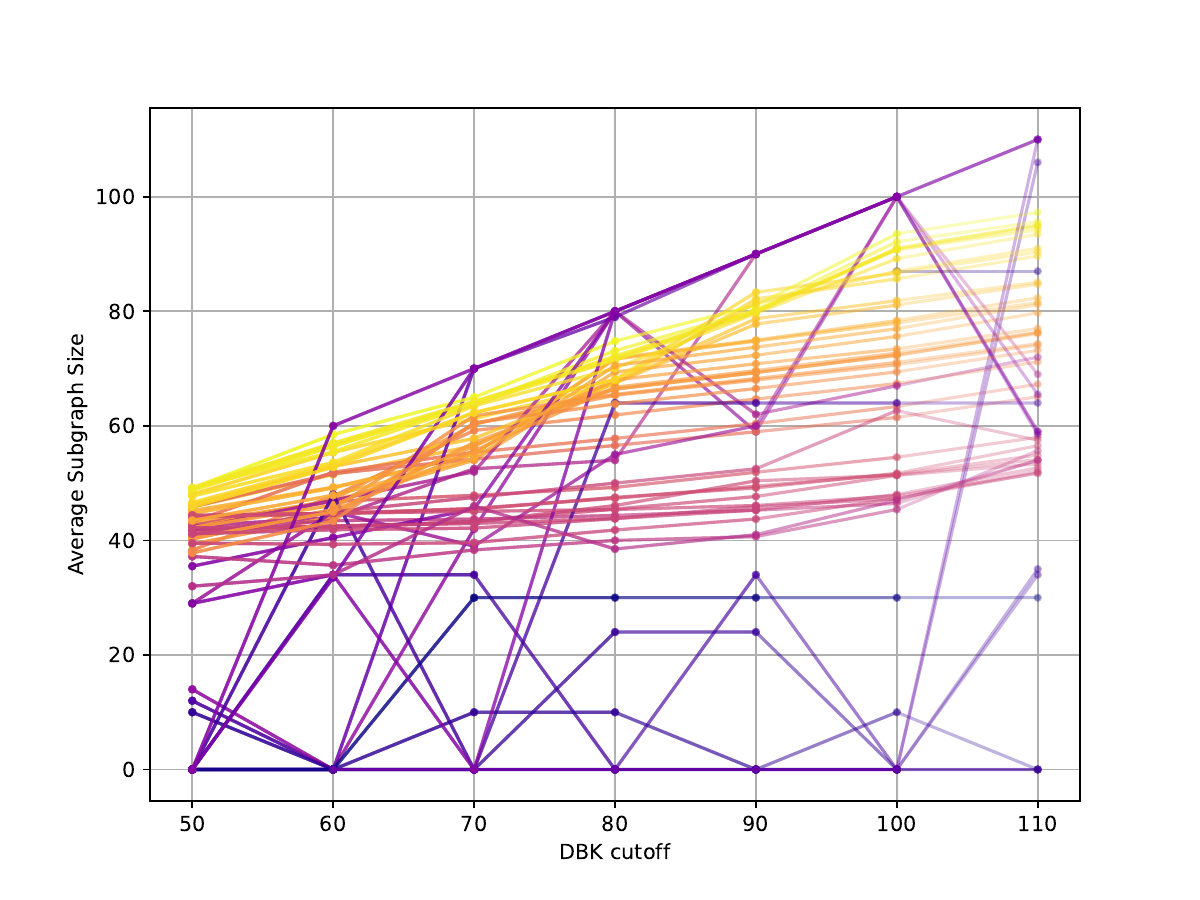}
    \includegraphics[width=0.07\textwidth]{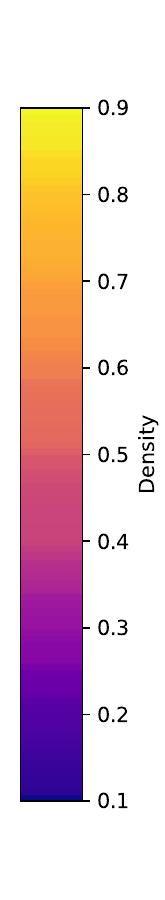}
    \caption{Number of subgraphs (top left), average subgraph density (top right), and average subgraph size (bottom) at each cutoff level of the DBK-fmc algorithm. Log scale on the y-axis of the top left plot. Input graph densities ranging from $0.1$ to $0.9$ (see color legend). \label{fig:DBK_decomposition_summary}}
\end{figure}

Figure~\ref{fig:DBK_decomposition_summary} shows the number of subgraphs, the average subgraph density, as well as the average size of the subgraphs generated at each cutoff level of the DBK-fmc algorithm. Note that the number of generated subgraphs can be zero (in particular, if a generated subgraph is a clique itself then it is not necessary to solve it, and therefore it not counted as a subgraph to be solved).

First, Figure~\ref{fig:DBK_decomposition_summary} (top left) shows that the number of generated subgraphs generally increases as the DBK cutoff decreases, as one should expect, especially for higher density graphs. At lower densities, the number of subgraphs is generally quite small (sometimes even zero) and does not greatly increase as the cutoff level decreases. An interesting case occurs for some high density graphs, whereby the trend of increasing generated subgraph counts actually reverses at a cutoff value of $60$. The precise reason why the reversal occurs at a cutoff value of $60$ is not clear, but the existence of a reversal is somewhat expected. This is due to the fact that two phenomena work against each other. A lower cutoff value causes the decomposition to run longer, thus producing more subgraphs. At the same time, an increase in the number of subgraphs results in all bounds working better, thus leading to more pruning.

Second, Figure~\ref{fig:DBK_decomposition_summary} (top right) shows that on average, as the cutoff decreases, the subgraph density increases in comparison to the density of the input graph. This is again a result of the design of the DBK algorithm, as the DBK algorithm preferentially removes lower degree nodes, both with the help of the k-core reduction and the low degree vertex removal when partitioning.

Third, Figure~\ref{fig:DBK_decomposition_summary} (bottom) shows the average subgraph size at each cutoff level. By construction of the DBK algorithm, the size of the subgraphs will be either equal to or smaller than the cutoff level. The latter case happens if either the bounds or the $k$-core reduction worked particularly well. As can be seen, the size of most subgraphs is indeed roughly equal to the cutoff for high densities, while subgraph sizes for lower densities vary widely.

\subsection{Solving the Maximum Clique problem on the subgraphs}
\label{sec:groundstates}
We are interested in exploring the accuracy of the  solutions  found by parallel quantum annealing on the leaf level of the DBK-fmc decomposition. First, for the test graphs of Section~\ref{sec:setup}, Figure~\ref{fig:parallel_embeddings} shows the disjoint minor-embeddings for different all-to-all problem graphs embedded on the Advantage System 4.1 Pegasus connectivity graph. We observe that a single anneal can find the maximum clique multiple times in a single anneal when using parallel embeddings. However, when calculating the ground state probability $p$ in eq.~\eqref{eq:TTS} over multiple samples, we only consider a binary indicator (maximum clique found at least once vs.\ not found) per sample. Note that, typically, single set of samples will contain a small number of optimal solutions (e.g., once or twice), if any.

\begin{figure}[h]
    \centering
    \includegraphics[width=0.69\textwidth]{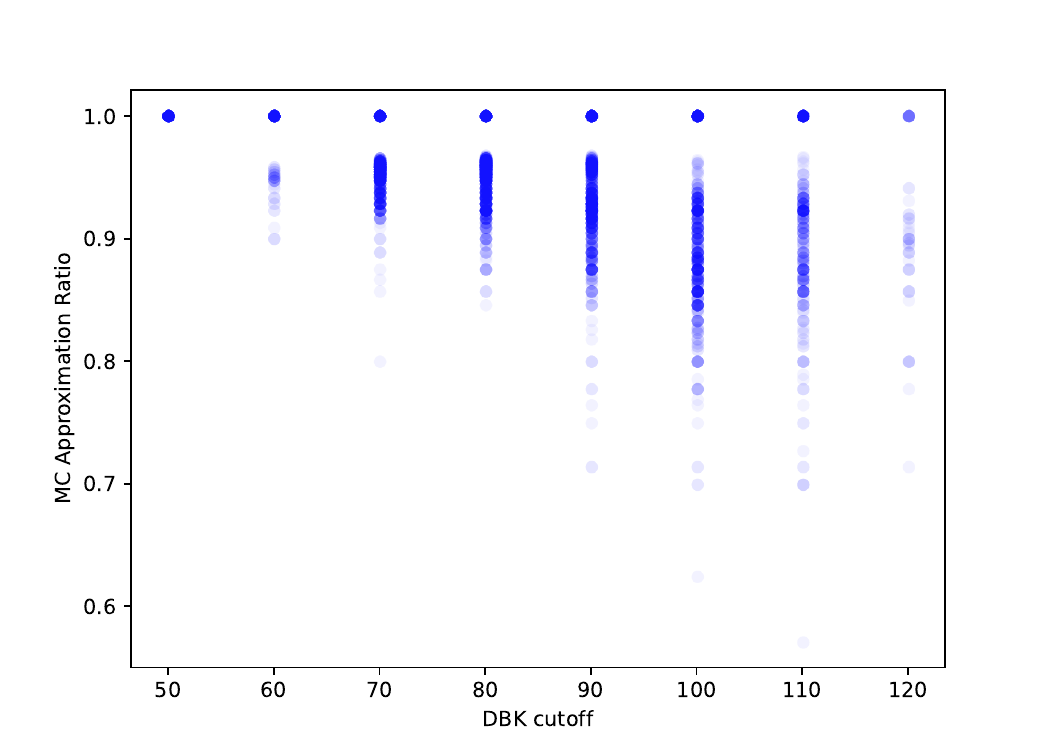}
    \caption{Scatter plot of the Maximum Clique approximation ratios across all subgraphs that were solved as a function of the DBK cutoff. Computed by taking the best Maximum Clique solution (unembedded with majority vote) out of the $1,000$ samples used for each subgraph. The data at a DBK cutoff of $120$ corresponds to the original input graphs.} \label{fig:approximation_ratios}
\end{figure}

Figure \ref{fig:approximation_ratios} shows the approximation ratios (specifically the approximation ratio of the largest clique found among the $1,000$ anneals) across all subgraphs generated during decomposition across the varying DBK cutoff values. Importantly, the approximation ratios consistently decrease as cutoff gets larger, and at a DBK cutoff of $50$, all of the subgraphs can be solved exactly. The difficult task is that for the original large graphs there is not a consistent way to know a-priori if the quantum annealer will find the maximum clique or not; this approximation ratio plot shows that decomposing the graph into smaller subgraphs increases the approximation ratio of the solutions found, up to being able to always find the maximum clique (at least for these $60$ random graphs). The approximation ratios in Figure \ref{fig:approximation_ratios} at a DBK cutoff of $120$ correspond to the non-parallel QA results from embedding a single 120 node clique onto the hardware and running the MC QUBO for each of the $60$ graphs on the hardware. Once the DBK decomposition is applied we see a gradual decrease in the failure rate to reach the optimal solution. 

\begin{figure}[h]
    \centering
    \includegraphics[width=0.49\textwidth]{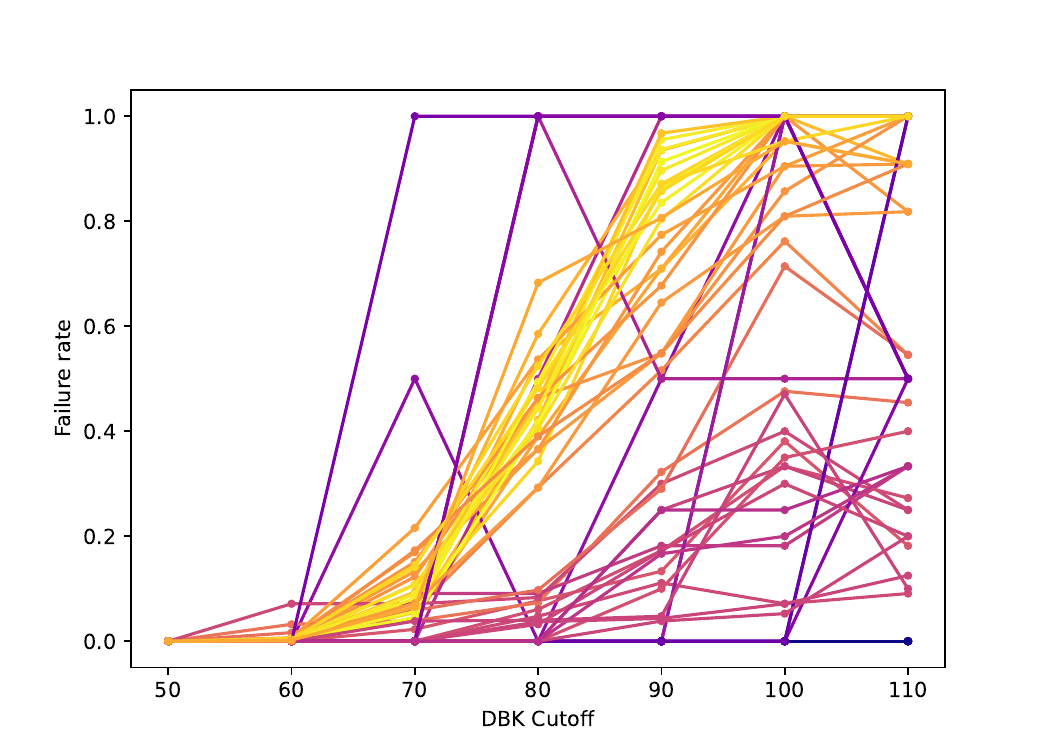}
    \includegraphics[width=0.49\textwidth]{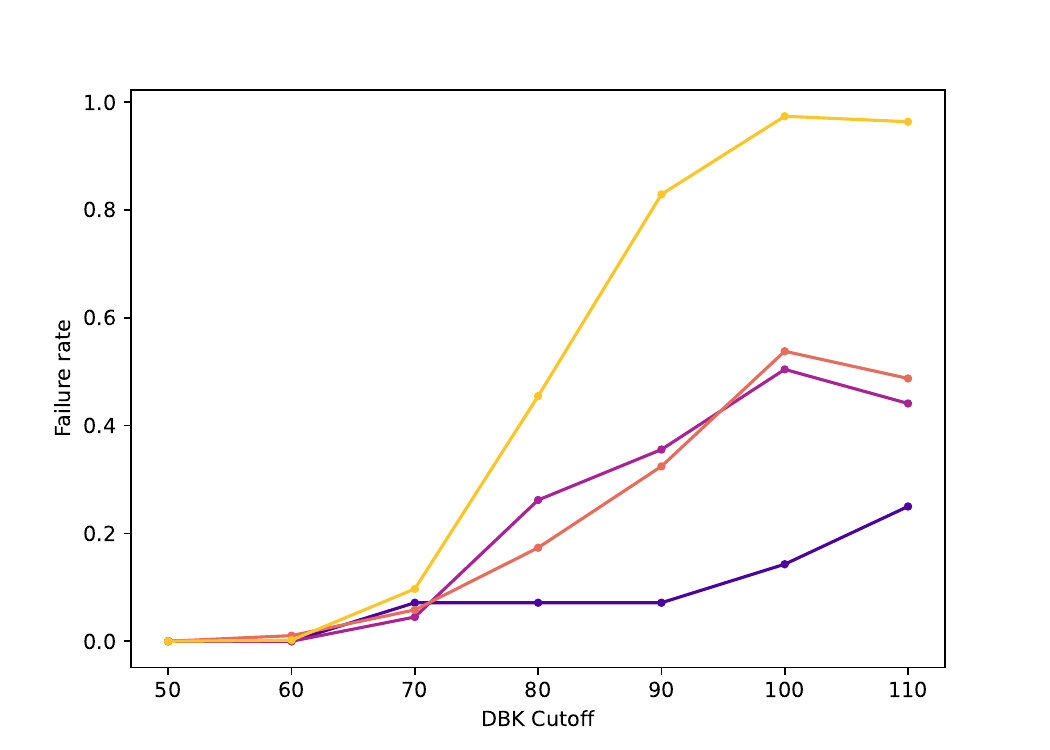}\\
    \includegraphics[width=0.5\textwidth]{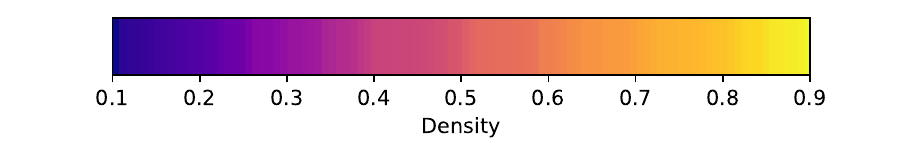}
    \caption{Left: Failure rate (failure to reach a ground state solution) as a function of the DBK cutoff values (after majority vote unembedding). Graph densities ranging from $0.1$ to $0.9$ (see color legend). Right: Averaged results for four groups of input graph densities ; $0.1-0.3$, $0.3-0.5$, $0.5-0.7$, and $0.7-0.9$ (the coloring corresponds to the median graph density of those ranges). \label{fig:failure_rate}}
\end{figure}

Next, Figure~\ref{fig:failure_rate} looks at the failure rate for finding ground state solutions over all $60$ test graphs as a function of the DBK cutoff value. We observe that for higher graph densities, ground state solutions are almost never found when using a DBK cutoff of $90$ or more, while for graphs decomposed down to sizes $70$ or less, the ground state of the subproblems is almost always found. As already observed in Section~\ref{sec:subgraphs}, the size and density of the subgraphs for lower input densities varies widely, resulting in either very easy or very hard MC problems for the quantum annealer. This causes the failure rate to be very volatile for lower input densities. 

\begin{figure}[h]
    \centering
    \includegraphics[width=0.32\textwidth]{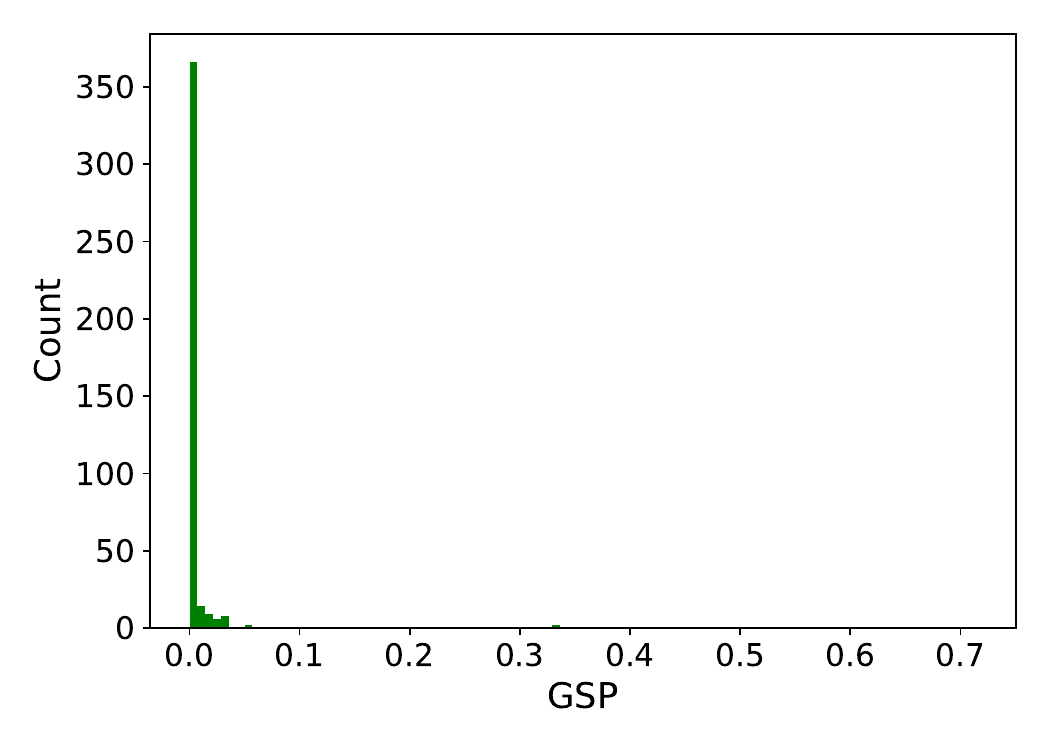}
    \includegraphics[width=0.32\textwidth]{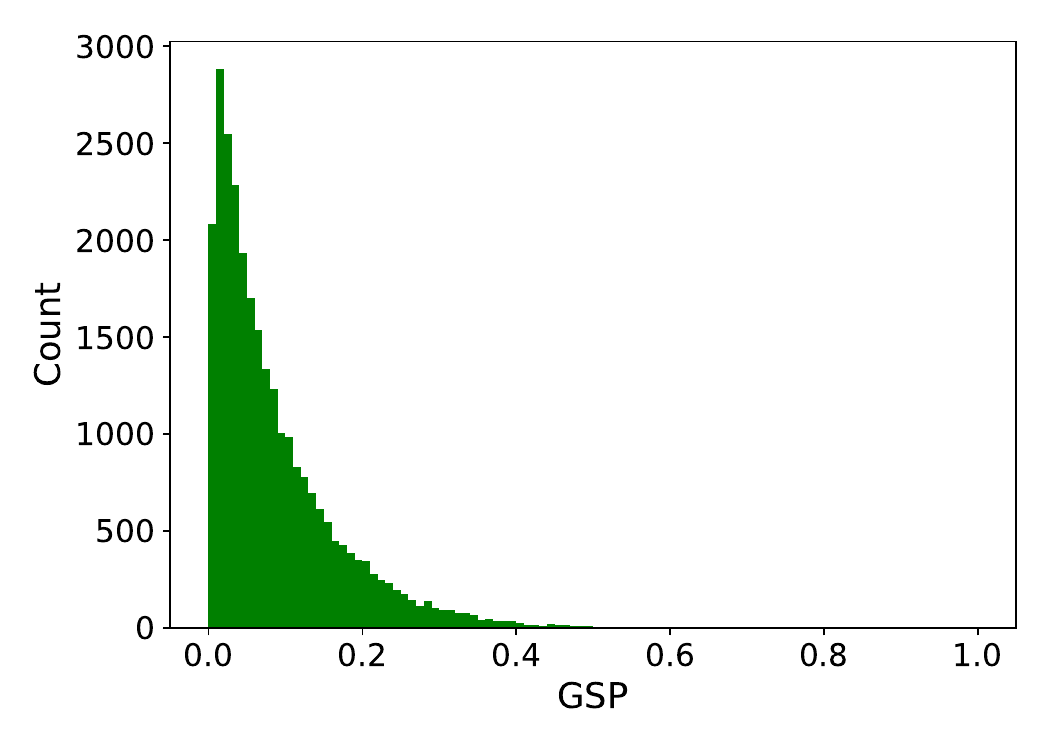}
    \includegraphics[width=0.32\textwidth]{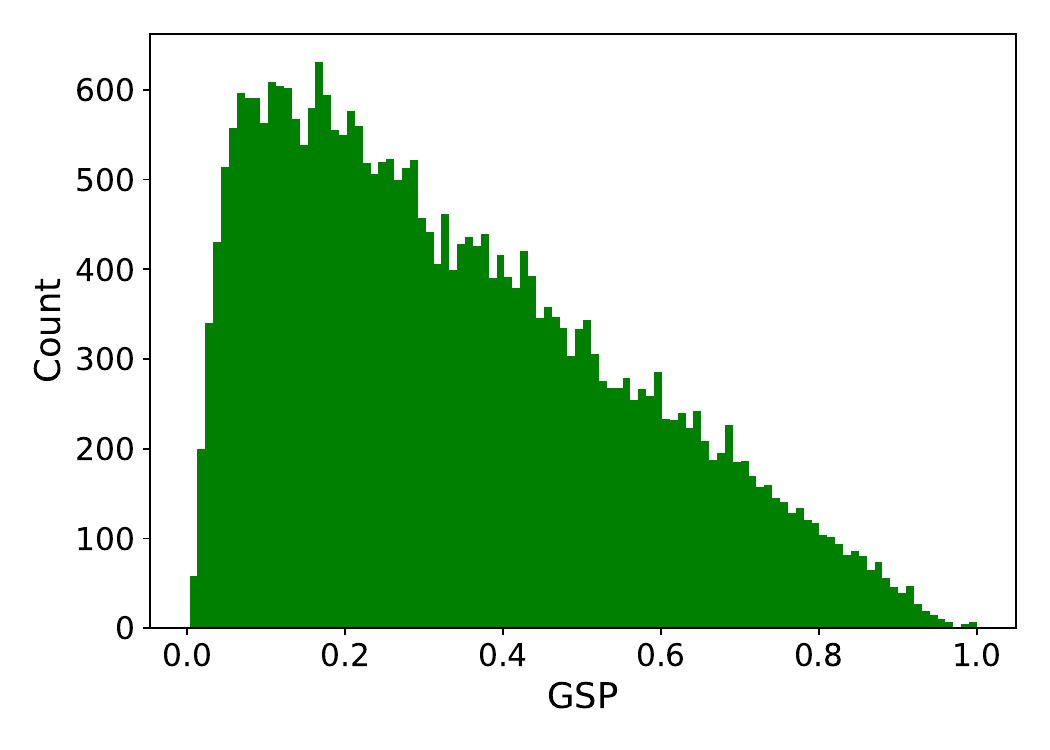}
    \caption{Histograms of the measured ground state probability of the Maximum Clique QUBO (after majority vote unembedding) across all subgraphs generated for a DBK cutoff value of $110$ (left), $60$ (middle), and $50$ (right).\label{fig:GSP_histograms}}
\end{figure}

Last, we visualize the number of subgraphs stratified by the proportion of samples that correctly found the optimal solution, referred to as the ground state probability (GSP) of the Maximum Clique QUBO(s), as a histogram. Figure~\ref{fig:GSP_histograms} shows that, as seen before, hardly any subproblem can be solved for a DBK cutoff of $110$. The lower the cutoff, the most subproblems can be solved to optimality by D-Wave. As seen in Figure~\ref{fig:failure_rate}, the GSP for a DBK cutoff of $50$ is always nonzero, meaning that when decomposing down to a subproblem size of $50$, all problems could be solved at least once to optimality in the $1000$ samples from the quantum annealer.

\subsection{Comparison of the DBK algorithm and FMC with the TTS measure}
\label{sec:comparison}
We compare the full DBK algorithm to the classical FMC solver using the TTS measure introduced in eq.~\eqref{eq:TTS}.

We note that the current noisy intermediate-scale quantum (NISQ) technology \cite{Preskill2018quantumcomputingin}, including D-Wave's quantum annealers, is not advanced enough and, hence, not competitive with classical computers on solving NP-hard optimization problems when dedicated algorithms are used. Regardless, we compare our algorithm against a highly optimized MC solver, the \textit{Fast Maximum Clique (FMC)} solver from \cite{fmc} in order to determine in which cases the quantum algorithm performance gets closer to the classical one.

\begin{figure}[h]
    \centering
    \includegraphics[width=0.49\textwidth]{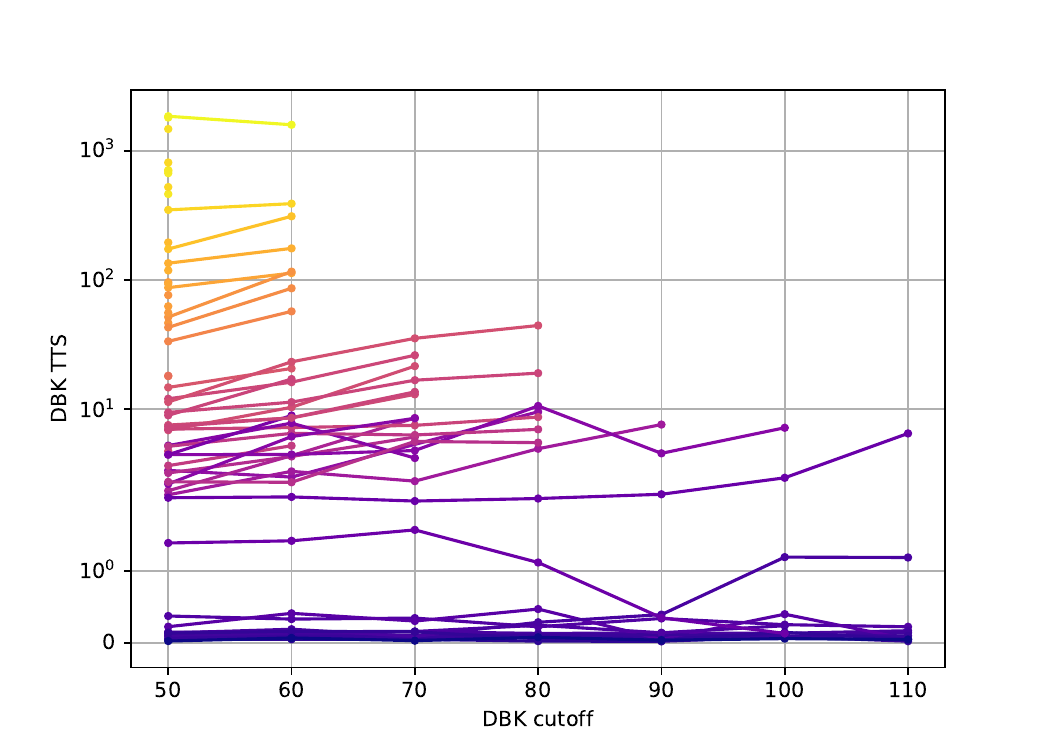}
    \includegraphics[width=0.49\textwidth]{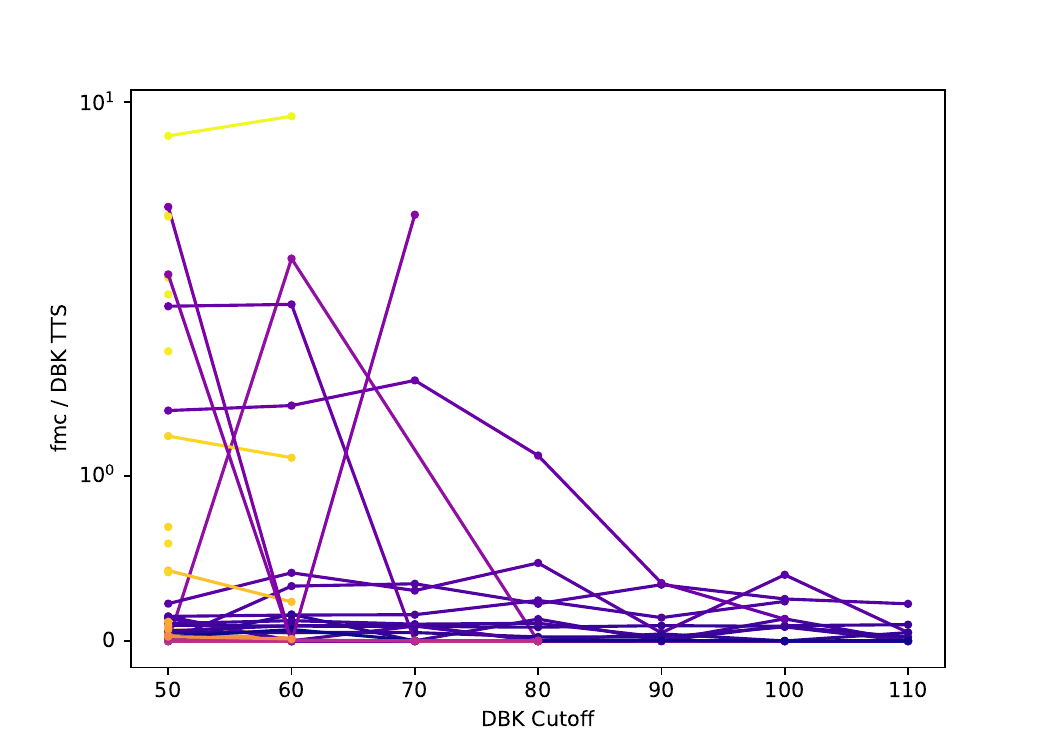}\\
    \includegraphics[width=0.49\textwidth]{figures/legend_horizontal.pdf}
    \caption{TTS as a function of the DBK cutoff for solving each of the $60$ test graphs. Each line represents a single graph being decomposed over different DBK cutoff values. Log scale on the y-axis. Graph densities ranging from $0.1$ to $0.9$ (see color legend). TTS time for the full DBK algorithm (including time for decomposition and unembedding) according to eq.~\ref{eq:TTS} (left), and FMC time divided by TTS time for the full DBK algorithm (right). \label{fig:TTS}}
\end{figure}

Figure~\ref{fig:TTS} shows several aspects of the comparison. First, Figure~\ref{fig:TTS} (left) shows the the TTS time of eq.~\eqref{eq:TTS} for the full DBK algorithm (including time for decomposition and unembedding) as a function of the DBK cutoff. We observe that the lower the density, the faster a test graph can be solved by the DBK algorithm. Moreover, high density graphs can only be solved when decomposing them to relatively low DBK cutoffs of $50$ to $60$, whereas lower density graphs can be decomposed at higher cutoffs as well. Importantly, at larger DBK cutoffs we have a lack of TTS values in the figure (especially at higher densities); this is because those TTS values could not be computed because for at least one subgraph $p=0$. This means that none of the $1000$ quantum annealing samples found the maximum clique at least once, which causes $p_i=0$ in eq.~\eqref{eq:TTS} which means a TTS value can not be computed. Thus another important aspect of Figure~\ref{fig:TTS} is showing at what densities and DBK cutoff values can \emph{all} subgraphs, that were generated by DBK-fmc, be solved by the quantum annealer in $1000$ samples. 

Second, Figure~\ref{fig:TTS} (right) shows the ratio of the classical FMC process time over the TTS time for the full DBK algorithm. Here, values above one indicate that FMC is slower than the quantum annealer sampling all subgraphs generated by the DBK-fmc algorithm. Indeed, we observe that the DBK-fmc algorithm with quantum annealing is superior to the entirely classical approach for high densities at low cutoff values of around $50$ or $60$, and computes maximum cliques up to two orders of magnitude faster than FMC. For either higher densities, or for higher cutoff values than $60$, the classical FMC algorithm is the better choice when solving the MC problem.

\subsection{Using D-Wave as a real-time subsolver for DBK}
\label{sec:experiments_real_time_DWave_solver}
In the previous experiments, we have emulated the procedure of executing DBK where the D-Wave quantum annealer is the subsolver. This was accomplished by exactly solving the problem graphs using FMC, and then determining what parameters (for example, annealing parameter and the DBK cutoff level) to use in order to consistently find the Maximum Clique.

In this section, we present results when using the quantum annealer, with the fixed disjoint clique embeddings for parallel quantum annealing, in the real time execution of DBK-pQA (with no classical solver). Based on our previous experiments (see Figure \ref{fig:GSP_histograms}), we would expect that with a DBK-pQA cutoff level of $50$ and using $1,000$ samples per subgraph, the Maximum Clique solution would be consistently found.

\begin{figure}[h]
    \centering
    \includegraphics[width=0.49\textwidth]{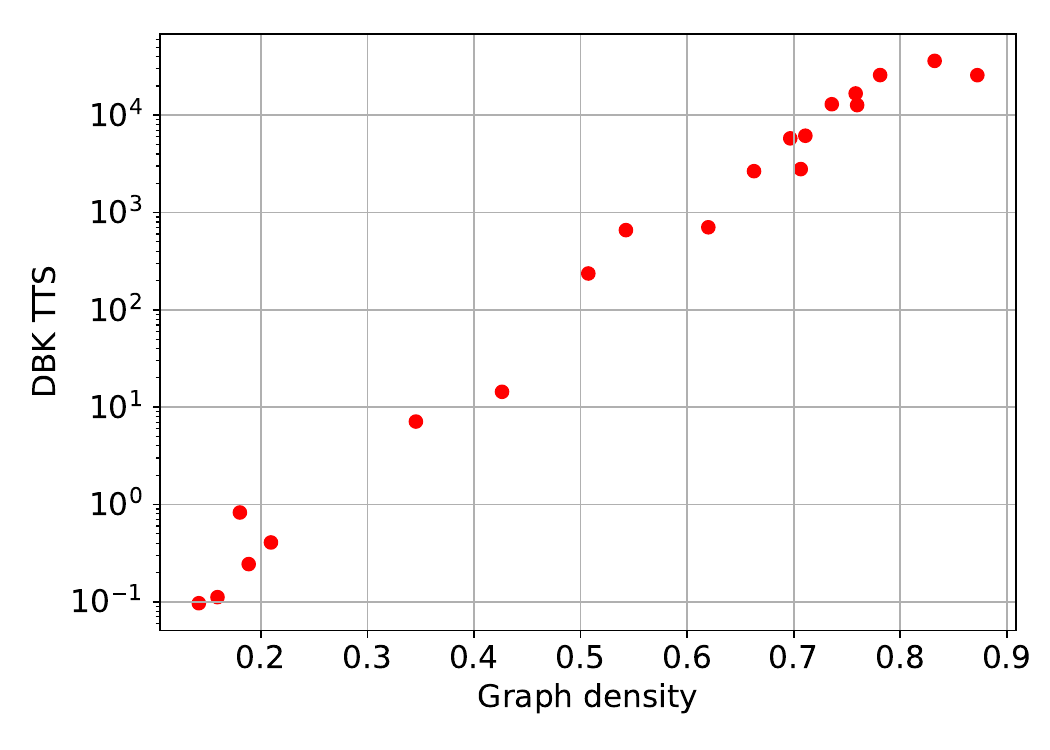}
    \caption{Scatter plot showing the DBK-pQA $TTS_{\text{fixed}}$ (y-axis) when using D-Wave as a real time solver in the DBK-pQA algorithm as a function of graph density. }
    \label{fig:DBK_TTS_real_time_DWave}
\end{figure}

In order to perform this experiment we ran the first $20$ random graphs of the $60$ generated for consistent experiments a total of $5$ times, while using the quantum annealer (with majority vote unembedding) as the subsolver. Using eq.~\eqref{eq:TTS_real_time_DBK} in order to compute the $TTS_{\text{fixed}}$, we can use the $5$ different runs for each graph in order to compute $p$. However, in all $100$ experiments, the algorithm found the maximum clique upon termination. Thus, $p=1$ in all cases, and therefore the time-to-solution was simply the average sum of the the used QPU time plus the used classical processing time. Figure~\ref{fig:DBK_TTS_real_time_DWave} shows the Time-to-Solution results.

The $TTS_{\text{fixed}}$ metric quantifies the success rate of the DBK-pQA algorithm, as opposed to the success rate of the individual samples from the quantum annealer. Therefore, for a large number of anneal samples (i.e., $1000$), we would expect the $TTS_{\text{fixed}}$ metric to be larger than the $TTS_{\text{opt}}$ metric. We observe this to be the case in Figure~\ref{fig:DBK_TTS_real_time_DWave}. Figure~\ref{fig:DBK_TTS_real_time_DWave} shows that the DBK-pQA TTS when using the quantum annealer as a real-time subsolver is exponential with respect to the graph density.

Most importantly, for all $20$ random graphs used in this experiment, DBK-pQA found an optimal solution to each original graph. This is despite the fact that D-Wave is a probabilistic sampler and DBK may generate up to $10,000$  subgraphs for each input graph that are each sent to D-Wave for finding a MC.

\section{Discussion}
\label{sec:discussion}
In this work, we consider solving the Maximum Clique problem using a combination of graph decomposition and parallel quantum annealing. To be precise, we base our solution on the DBK algorithm \cite{Pelofske2019mc} to decompose an arbitrary input graph to subgraphs of any pre-specified cutoff level sizes. We then employ the D-Wave Advantage System 4.1, a quantum annealer manufactured by D-Wave Systems, Inc., to solve the subproblems generated during the decomposition. In order to best leverage the capabilities of the D-Wave annealer, we embed and solve several of the generated subproblems simultaneously, an idea previously introduced under the name of parallel quantum annealing in the literature \cite{Pelofske2022parallel}. Therefore, this work shows the end-to-end process required to solve large maximum clique problems to (near) optimality using a quantum annealer.

Using the DBK algorithm in connection with D-Wave Advantage System 4.1, we are able to sample ground state solutions of MC problems of up to $120$ vertices, and additionally employ parallel quantum annealing to speed up the computations. We demonstrate that in some cases, the resulting algorithm can compute maximum cliques in high density graphs up to around two to three orders of magnitude faster than a classical solver. Current NISQ annealing devices are not necessarily expected to outperform classical algorithms, although a scaling advantage using quantum annealing has been demonstrated in \cite{King_2021}. The experiment reported in \cite{King_2021} considers the simulation of geometrically frustrated magnets (which reduces to a 0-1 integer programming problem on a given geometric lattice) using quantum annealing, and demonstrates that a quantum annealing processor can provide a computational advantage over the classical counterpart, path-integral Monte Carlo (PIMC).

This work leaves scope for further research avenues:
\begin{enumerate}
    \item The methodology of this article (the exact graph decomposition in connection with multiple embeddings) can be applied to other NP-hard problems such as the minimum vertex cover problem. 
    \item The performance of the quantum annealing backend can be improved by using further tunable features such as h-gain schedules, anneal offsets, flux bias offsets, and spin reversals.
    \item It remains to be analyzed how different minor embeddings solve the same QUBO when they are all used in the same annealing cycle. In particular, it is unclear if some embeddings contribute more than others to finding the ground state solutions. If so, what characteristics of those embeddings cause them to perform better than the other embeddings? Additionally, the spatial and temporal correlations with regard to which embeddings are finding maximum cliques remain to be investigated.
    \item Do minor embeddings that have the same connectivity perform the same or differently during the same anneal(s)? In other words, if all of the disjoint embeddings used in the parallel quantum annealing method are exactly the same, just acting on different qubits, are the ground states found across the different embeddings with (nearly) equal probabilities?
    \item Utilizing an efficient algorithm to compute the minimum number of samples required in order to obtain an optimal TTS (as opposed to using a fixed number of samples) would significantly reduce the real time computation required to run DBK with a quantum annealer. 
\end{enumerate}

\section{Data availability}
Python implementation and data for parallel quantum annealing and the DBK graph decomposition algorithm can be found on Github \cite{python}.

\section{Acknowledgments}
\label{sec:acknowledgments}
This work was supported by the U.S. Department of Energy through the Los Alamos National Laboratory. Los Alamos National Laboratory is operated by Triad National Security, LLC, for the National Nuclear Security Administration of U.S. Department of Energy (Contract No. 89233218CNA000001). The research presented in this article was supported by the Laboratory Directed Research and Development program of Los Alamos National Laboratory under project number 20220656ER. The work of Hristo Djidjev has been  also partially supported by the Grant No BG05M2OP001-1.001-0003, financed by the Science and Education for Smart Growth 
Operational Program (2014-2020) and co-financed by the European Union through the European structural and Investment funds.

This research used resources provided by the Darwin testbed at Los Alamos National Laboratory (LANL), which is funded by the Computational Systems and Software Environments subprogram of LANL's Advanced Simulation and Computing program (NNSA/DOE). This research used resources provided by the Los Alamos National Laboratory Institutional Computing Program. Research presented in this article was supported by the NNSA's Advanced Simulation and Computing Beyond Moore's Law Program at Los Alamos National Laboratory. 

\noindent
LA-UR-22-24790


\setlength\bibitemsep{0pt}
\printbibliography

\end{document}